\documentclass[apj]{emulateapj}

\usepackage[utf8]{inputenc}
\usepackage{graphicx}
\usepackage{apjfonts}
\usepackage{url}
\usepackage{amsmath}
\usepackage{epstopdf}
\usepackage{tablefootnote}
\usepackage{threeparttable}
\usepackage{hyperref}
\usepackage{amssymb}
\usepackage{amsopn}
\newcommand{\angstrom}{\textup{\AA}}

\begin{document}
%Titre
%\maketitle
\title{\textsc{BANYAN. VIII. New low-mass stars and brown dwarfs with candidate circumstellar disks.}}

%Auteur
\author{Anne Boucher$^1$}
\author{David Lafreni\`ere$^1$}
\author{Jonathan Gagn\'e$^{2,3}$}
\author{Lison Malo$^{4}$}
\author{Jacqueline K. Faherty$^{2,5,6}$}
\author{Ren\'e Doyon$^1$}
\author{Christine H. Chen$^{7}$}

\affiliation{$^1$ Institut de Recherche sur les Exoplan\`etes (iREx), Universit\'e de Montr\'eal, D\'epartement de Physique, C.P. 6128 Succ. Centre-ville, Montr\'eal, \\ QC H3C 3J7, Canada. \texttt{boucher@astro.umontreal.ca}}
\affiliation{$^2$ Department of Terrestrial Magnetism, Carnegie Institution for Science, 5241 Broad Branch Road NW, Washington, DC 20015, USA}
\affiliation{$^3$ NASA Sagan Fellow}
\affiliation{$^4$ Canada-France-Hawaii Telescope, 65-1238 Mamalahoa Hwy, Kamuela, HI 96743, USA}
\affiliation{$^5$ Department of Astrophysics, American Museum of Natural History, Central Park West
at 79th Street, New York, NY 10034}
\affiliation{$^6$ NASA Hubble Fellow}
\affiliation{$^7$ Space Telescope Science Institute, 3700 San Martin Drive, Baltimore, MD 21218, USA}

\begin{abstract}
We present the results of a search for new circumstellar disks around low-mass stars and brown dwarfs with spectral types >K5 that are confirmed or candidate members of nearby young moving groups. Our search input sample was drawn from the BANYAN surveys of Malo et al. and Gagn\'e et al. \emph{Two-Micron All-Sky Survey} and \emph{Wide-field Infrared Survey Explorer} data were used to detect near- to mid-infrared excesses that would reveal the presence of circumstellar disks. A total of 13 targets with convincing excesses were identified: four are new and nine were already known in the literature. The new candidates are 2MASS~J05010082--4337102 (M4.5), J08561384--1342242 (M8$\,\gamma$), J12474428--3816464 (M9$\,\gamma$) and J02265658--5327032 (L0$\,\delta$), and are candidate members of the TW Hya ($\sim10\pm 3\,$Myr), Columba ($\sim 42^{+6}_{-4}\,$Myr) and Tucana-Horologium ($\sim 45\pm 4\,$Myr) associations, with masses of $120$ and 13--18\,$M_{\mathrm{Jup}}$. The M8--L0 objects in Columba and Tucana-Horologium are potentially among the first substellar disk systems aged $\sim 40\,$Myr. Estimates of the new candidates' mean disk temperatures and fractional luminosities are in the ranges $\sim 135-520\,$K and $0.021-0.15$, respectively. New optical spectroscopy of J0501--4337 reveals strong H$\alpha$ emission, possibly indicating ongoing accretion, provides a detection of lithium absorption and a radial velocity measurement that is consistent with a membership to Columba. We also present a near-infrared spectrum of J0226--5327 that reveals Paschen $\beta$ emission and shows signs of low surface gravity, consistent with accretion from a disk and a young age.\\
\textit{Subject headings: } Circumstellar matter, Brown dwarfs, Stars: Low-mass, Protoplanetary disks, Infrared: Stars
\end{abstract}

\section{Introduction}
\label{sec:Intro}
The presence of debris disks is a signpost of a past planetary formation \citep{Zuckerman2004P, Moor2011, Raymond2011}, whereas the presence of protoplanetary or transition disks can be a sign of ongoing planetary formation \citep{Espaillat2010,Kraus2011}. Searching for circumstellar disks is thus an interesting way to identify new exoplanets and study their formation. Disks are usually initially detected from the presence of an excess of infrared (IR) radiation compared to that expected from the star alone. Such an excess is attributed to the thermal emission of the dust in the disk that is warmed up by its host star. The amount of excess emission can vary greatly from one system to another, and reveals different types of disks or different stages in their evolution. Primordial disks are massive ($>10\,M_{\oplus}$; \citealt{Wyatt2008}), optically thick, rich in gas and accreting onto their host star, and produce an important excess emission from the near-infrared (NIR) and on-wards; they are found around the youngest systems ($<10\,$Myr; \citealt{Williams2011}). Dusty debris disks are much more tenuous ($<1\,M_{\oplus}$; \citealt{Wyatt2008}), colder and gas poor; they are thought to be the remnants of a planetary system formation and are constantly replenished by planetesimal or asteroid collisions, which allows them to be sustained for much longer, up to 100 Myr \citep{Zuckerman2004P,Moor2006}. Another difference between primordial and debris disks can be found int their dust grain sizes: those of the former are smaller and more amorphous (indicating a lesser amount of grain growth and thermal processing). Since debris disks are colder ($\sim$~10--100\,K), their weaker excesses are mostly observable in the mid-infrared (MIR) and beyond ($>20\,\mu$m; \citealt{Rhee2007,Schneider2012a,Liu2014}). Transition disks (TD) are intermediate cases: they feature an inner region devoid of warm dust, and an outer region that still contains a significant amount of dust. Transition disks display an excess starting at about $10\,\mu$m \citep{Strom1989, Espaillat2010}, and may still contain some amount of gas.

Circumstellar disks have been found and studied around many types of stars, spanning early spectral types (e.g., A stars like the well-known Vega (A0\,V), $\beta\,$Pictoris (A6\,V) and Fomalhaut (A4\,V); \citealt{Aumann1985,Hobbs1985}) to late-type stars (e.g., K and M classes; \citealt{Avenhaus2012,Kraus2014,Theissen2014}), and down to the substellar and planetary-mass regimes \citep{Luhman2005b,Luhman2005a,Bayo2012,Joergens2013}. However, the number of disks that are currently known around very low-mass stars and substellar objects remains relatively small, particularly for debris and transition disks. Improving the current sample of such systems would be valuable for many reasons. Since the hosts of such disks are more likely to host planets and have a lower intrinsic luminosity, they would be important targets for direct-imaging searches of exoplanets. In addition, the occurrence and properties of disks in these low-mass systems -- with masses near the lower limit of star formation -- would offer some unique insights into the star, planet and disk formation and evolution mechanisms.%.%For instance, because of their higher likelihood of hosting planets and the lower intrinsic luminosity of their host star

In the last few years, there has been a lot of effort to identify new young low-mass stars and brown dwarfs (BDs) in the solar neighborhood (e.g. \citealt{Rodriguez2011,Malo2013,BANYANV}), providing many good candidates for the search of new disks in the low-mass regime. The proximity of these systems also makes it easier to spatially resolve and image their disks. We have thus carried out a search for excess infrared emission around several newly identified very low-mass stars and BDs in nearby young moving groups (YMGs) and associations ($\lesssim250\,$Myr; \citealt{Bell2015}). The stars within a given group formed at the same time, most probably from the same molecular cloud, and are found with similar galactic positions and space velocities \citep{Zuckerman2004}.  The youngest and closest ($\lesssim 100\,$pc) associations that we know of are the following: the TW Hya association (TWA; $7-13\,$Myr; \citealt{delaReza1989,Kastner1997,Zuckerman2004,Weinberger2013}), $\beta$ Pictoris ($\beta$PMG; $21-27\,$Myr; \citealt{Zuckerman2001Bpic, Malo2014b, Binks2014}), Tucana-Horologium (THA; $41-49\,$Myr; \citealt{Torres2000, Zuckerman2000, Zuckerman2001Tuc, Kraus2014}), Carina (CAR; $38-56\,$Myr; \citealt{Torres2008}), Columba (COL; $38-48\,$Myr; \citealt{Torres2008}), Argus (ARG; $61-88\,$Myr; \citealt{Makarov2000}) and AB Doradus (ABDMG; $130-200\,$Myr; \citealt{Zuckerman2004ABDor, Luhman2005ABDor, Barenfeld2013}). %This is why YMGs are very useful to constrain the age of a system. %These are the associations targeted by the BANYAN surveys (ref????), from which our target sample was drawn (see Section \ref{sec:Sample}).***%(excluding the star-forming regions)

In this paper, we present the results of our search for new circumstellar disks around young low-mass stars and brown dwarfs, which led to the identification of 4 systems with newly detected significant excesses. In Section \ref{sec:Sample}, we present our target sample and our selection criteria. We detail our method to identify the infrared excesses in Section \ref{sec:Method}. In Section \ref{sec:Obs} we present follow-up and complementary observations, and the results of our search are presented in Section \ref{sec:Results}. Lastly, we discuss these results and our current conclusions in Section \ref{sec:Discuss}.

\section{Target Sample}
\label{sec:Sample}

We used the BASS survey catalog of \citealt{BANYANVII} (BANYAN All-Sky Survey; \citealt{BANYANV}), the candidate list from \citet{Malo2013} and the bona fide list from \citet{Malo2014b} of nearby moving group members or candidate members (see the BANYAN article series) as a starting point to build our target sample. The BASS catalog includes stars with spectral types $\geq$ M5 while the objects in \cite{Malo2013,Malo2014b} include earlier-type stars (K5\,V--M5\,V). Their ages range from $\sim 10$--250\,Myr, depending on the age of the moving group they are in. We started with a total of $\sim 1600$ targets from all those lists, and required that the stars have data both in 2MASS \citep{2MASS} and WISE \citep{WISE}. When a star could not be found in one or the other catalog, we used the cross-matched targets from \cite{BANYANV}. We also required our sample to have good detections in the $W3$ and $W4$ bands (i.e., no upper limits) and not to be flagged by WISE as contaminated and/or extended from their primary data set (profile-fit photometry). Lastly, we required a signal-to-noise ratio (SNR) $>3$ in the $W4$ band. After applying these selection criteria, 585 objects remained: 324 from the BASS catalog and 261 from the lists of Malo et al.%(see \citealt{Bell2015} for the most recent group ages)

\section{Method}
\label{sec:Method}
\subsection{Initial Excess Detection}
\label{subsec:InitDetect}
In order to determine whether or not a star had an infrared excess, we first fit synthetic spectra (BT-Settl; \citealt{Allard2013}) to the observed photometric data points by using a least-squares approach over the following photometric bands: I (when available from the DENIS $3^{rd}$ release; 2005) and 2MASS $J, H$ \& $K_s$. The 2MASS, DENIS (and WISE, for the analysis done later) magnitudes were converted to flux densities ($f_{\lambda}$) using the zero points given by \cite{Cohen2003}, \cite{DENIS} and \citealt{Jarrett2011}, respectively, and synthetic flux densities ($f_{\lambda,\mathrm{mod}}$) across the corresponding bandpasses were computed from the synthetic spectra using the relative spectral response curves given in the same references. For a given object, the predicted photospheric flux densities were defined as $f_{\lambda,\mathrm{pred}}=\left \langle af_{\lambda,\mathrm{mod}} \right \rangle$, where the mean is calculated over the synthetic flux densities of all the models that produced a fit with $\chi^2_{\mathrm{red}}$ lower than the 68\% confidence level limit according to the $\chi^2$ formalism, and where $a$ is a numerical factor explained below. When there were fewer than 5 models that satisfied this criterion, we only kept the 5 models with the smallest $\chi^2_{\mathrm{red}}$ values. We defined $\sigma_{\lambda,\mathrm{pred}}$ as the standard deviation of the flux densities from this set of best models.

We only considered the range of BT-Settl models that are most appropriate to our target sample for the spectral fitting of the photometry (late-type stars and brown dwarfs: mid-K through early T), namely $T_{\mathrm{eff}} \leq 5000\,$K, $\log g$ $> 3.5$, and with solar metallicity. Furthermore, all of the targets from \cite{BANYANVII} and from \cite{Malo2013,Malo2014b} were already assigned a spectral type estimate. We took advantage of this and, based on those estimates, assigned an effective temperature ($T_{\mathrm{SpType}}$) to each star, using the spectral type--$T_{\mathrm{eff}}$ relations of \cite{Stephens2009} for the M6 to T1 types and those of \cite{Pecaut2013} for earlier-type stars (<M5). Then, only the models with $T_{\mathrm{eff}}$ within the range of $T_{\mathrm{SpType}}$ $-500$ to $+100\,$K were considered in the spectral fitting; this was useful to avoid any secondary minima during the fitting. We went down to $-500\,$K because there are indications that young stars and brown dwarfs (and also stars with lower surface gravity, which usually indicates a young age) have cooler effective temperatures when compared to field stars and brown dwarfs with the same spectral types \citep{Filippazzo2015,BANYANVII,Metchev2006,Barman2011a,Barman2011b,Males2014,Faherty2012,Liu2013b}.

We calculated the reduced $\chi^2$ for every model using the following equation:
\begin{equation}
\label{eq:chi2}
\chi^2_{\mathrm{red}} = \frac{1}{K} \sum_{\mathrm{bands}}  \frac{\left( f_{\lambda}-af_{\lambda,\mathrm{mod}}\right) ^2}{\sigma_{\lambda}^2},
\end{equation}
where $K$ is the number of degrees of freedom and $a$ is the fitted solid angle ($=\pi R^2/D^2$ ; $R$ is the modeled star radius and $D$ its hypothetical distance), which minimizes the $\chi^2$ and can be found analytically. Then, the best models that minimize $\chi^2$ were selected.

In order to characterize the excess in the WISE bands ($W1$ to $W4$), we calculated its significance level and considered the uncertainty that could come from the fit itself, using: 
\begin{equation}
\label{eq:S}
S = \frac{f_{\lambda}-f_{\lambda,\mathrm{pred}}}{\sqrt{\sigma_{\lambda}^2+\sigma_{\lambda,\mathrm{pred}}^2}},
\end{equation}
where $f_{\lambda}$ is the observed photometric flux density of a given band ($\sigma_{\lambda}$ is the error of $f_{\lambda}$).% while $f_{\lambda,\mathrm{pred}}$ and $\sigma_{\lambda,\mathrm{pred}}$ are described above. 

\begin{figure}
\centering
    \includegraphics[scale=0.41]{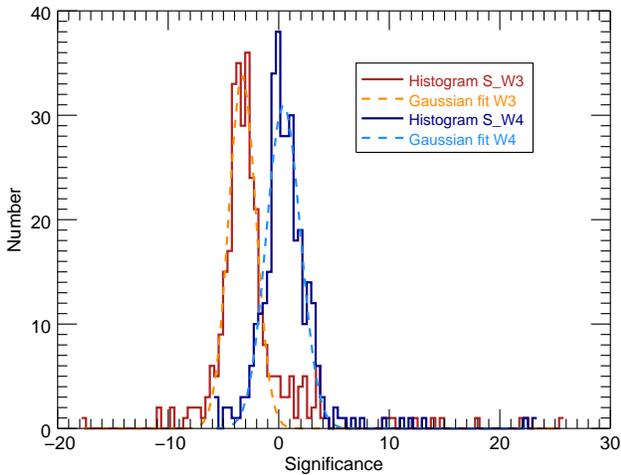}
\caption{Histogram of the significance distribution of the excesses in the WISE $W3$ (red) and $W4$ bands (blue), as well as their associated Gaussian fit (in orange and light blue, respectively). }
\label{fig:histo_S}
\end{figure}

The distributions of $S$($W3$) and $S$($W4$) for our full target list is shown in Figure~\ref{fig:histo_S}. The bulk of both distributions are well fit by a Gaussian, as expected, with characteristic widths ($\sigma$) of 1.2 in $W3$ and 1.5 in $W4$, respectively, indicating that our estimates for the uncertainty of the expected model fluxes are reasonable.
The distribution of $S$($W4$) is centered close to 0, as expected if the models correctly match the observed flux, but the distribution of $S$($W3$) is significantly shifted towards negative values ($\sim\,-2.9$), indicating that the models systematically over-predict the $W3$ flux. The reason for this systematic is unclear, but it has no significant bearing on our study, apart from possibly missing some sources with smaller excesses.

We identified candidates by selecting those with a significant infrared excess when compared to the flux expected from the best model fits in at least the $W3$ or $W4$ WISE bands. We then applied the two following selection criteria: 

(A) The excess must have a significance level of at least $3\,\sigma$, 
\begin{equation}
S \geq 3,
\end{equation}
as defined in Equation~\eqref{eq:S}.

(B) The observed flux must be larger than the synthetic flux by 60\% in $W3$ and by a factor of 3 to 10 in $W4$:
\begin{eqnarray}
p_{W3} = \frac{f_{W3}}{f_{W3,\mathrm{pred}}} \geq 1.5, \\
p_{W4} = \frac{f_{W4}}{f_{W4,\mathrm{pred}}} \geq 3-10. 
\end{eqnarray}
The values for those thresholds have been determined subjectively by looking at the typical values of $p_{W3}$ and $p_{W4}$ for a sample of non-excess sources and comparing them to a sample of what seemed to be significant excesses. If there was a $3\sigma$ detection in the $W3$ band, then $p_{W4}$ could go as low as 3. If not, then $p_{W4}$ had to be at least equal to 10 or higher to be considered any further. %We tried to be as inclusive as possible: 

Multiple fits were then performed on each target: the model spectra were first fitted to the $J, H, K_s$ (and $I$, when available) bands (to only fit the star) and then the WISE bands (where an excess can be expected) were progressively added to the fit. This method was useful to determine whether an excess was real or if it could be explained to a large extent by a normal photosphere with different values of $T_{\mathrm{eff}}$ and $\log g$. When these ``multi-fits'', using all sets of bands, all converged on the same model, it was interpreted as a sign that the excess is most likely real and that the model represented the stellar spectral energy distribution (SED) well enough. In these cases, the initial fits (i.e. with ($I$,) $J, H$ and $K_s$ only) were adopted.

We applied this method to two data sets of WISE photometry: the \textit{instrumental profile-fit photometry magnitude} and the \textit{instrumental standard aperture magnitude with aperture correction applied} data sets. We observed that the $W4$ band (and sometimes $W3$) often had a good detection in the \textit{profile-fit} set while only an upper limit or a contaminated measurement was available in the \textit{aperture} photometry. We have thus used both sets and applied the same selection criteria in both cases.

If 3 or 4 out of the 4 criteria mentioned above were satisfied for a given fit, then the target was identified as a disk candidate. If only 2/4 criteria were satisfied, the target was flagged as ``ambiguous'' and we kept it for further examination. When one or none of the criteria were satisfied, the target was completely rejected. Since this was done on two WISE data sets, we required that a given target be selected as a good or ambiguous disk candidate in both sets for it to be preserved in the final sample. 

\subsection{Detailed Verification of Candidate Stars with Excess}
\label{subsec:Verif}
Once the targets with the most significant infrared excesses were identified, additional in-depth verifications were made to confirm that the detections were due to a real excess and not to false-positives and/or contaminated data. First, a visual inspection of the $W1$ to $W4$ WISE images was done. It was verified that a signal was present in the image, that it looked distinct from the noise at the source position and that there was no other source within the largest WISE aperture ($16.5''$, associated with the $W4$ channel). Secondly, the point-spread function (PSF) shape was inspected to ensure that it was round and did not display an extended shape or a surrounding haze. A PSF shape that is not round or that is extended could indicate some kind of artefact affecting the photometry of the source. When no clear point-like signal was present in any one of the 2 bands, the target was discarded: we required a visual detection in \emph{both} channels. The same verification was done with the 2MASS and the Digitized Sky-Survey (DSS) images. The smaller angular resolution of 2MASS and DSS ($\sim 2''$ and $\sim 1''$, respectively) allowed for a better rejection of contaminants due to background stars. Any target that displayed signs of contamination or an extended shape was discarded. This step removed most of the contaminants after the initial excess detection. 

On a few occasions during the verification mentioned above, it was noticed that a star with an apparently significant $W4$ excess (in either photometry sets) seemed to contain no flux at the source position in the image except for noise. For this reason, the WISE magnitudes were re-calculated for all of the disk candidates to verify that their reported magnitudes were not similarly affected by this apparent problem. Standard aperture photometry was used with an aperture radius set to the median full-width half-max (FWHM) of a sample of 20 bright and isolated WISE stars with SNR >10 in the $W4$ band. The appropriate aperture correction factor and its uncertainty were calculated from the same set of stars. In order to estimate the uncertainty on these measurements, the flux density was measured using multiple circular apertures at random positions around the target. It was required that the random apertures do not overlap and do not contain any real source. The standard deviation of the fluxes that were measured in these random apertures was taken as the flux measurements uncertainty, and was translated to an uncertainty on its magnitude. %  of data, the \textit{profile-fit} or the \textit{aperture} one

A good agreement was obtained between these re-calculations and the cataloged WISE magnitudes for the disk candidates that respected all selection criteria mentioned above. In most cases, the differences were smaller than $1\,\sigma$, while the worst cases differed by up to $2\,\sigma$. Thus, even when considering these disparities, all excess detections remained significant at more than the $3\,\sigma$ level. For the remainder of this work, only the \textit{profile-fit} data set from the WISE catalog is considered. 

\subsection{Blackbody Fit}
\label{subsec:BBfit}

The mean temperatures of the circumstellar disks causing the IR excesses detected in Section~\ref{subsec:InitDetect} can be estimated by fitting a blackbody spectrum to the IR photometry. This was first performed using the best-fit model spectra previously identified and all available photometry. However, it was observed that the 2MASS data points pulled the blackbody spectrum fit to higher temperature to compensate for small differences at short wavelengths, not capturing well the excess at $W3$ and $W4$. Because of this, the excess $E = f_{\lambda}-f_{\lambda,\mathrm{pred}}$ was directly fitted over the $W1$ to $W4$ bands only. The error on the excess was defined as $\sigma_E = \sqrt{\sigma_{\lambda}^2+\sigma_{\lambda,\mathrm{pred}}^2}$. The $\chi^2$ as defined in Equation~\eqref{eq:chi2} was then minimized, but using $B(T_{\mathrm{disk}})$ -- the Planck emission function of a blackbody at a temperature $T_{\mathrm{disk}}$ -- instead of $f_{\lambda,\mathrm{mod}}$. The outputs of this method were $a$, the flux scaling factor, and $T_{\mathrm{disk}}$, the estimated mean disk temperature. 

A Monte Carlo approach was used to estimate errors on the best fitting parameters. Random Gaussian noise was added to the data according to $\sigma_E$ in $E$ and $T_{\mathrm{disk}}$ and $a$ were evaluated for every synthetic data set. The 68\% confidence interval of the resulting distribution was taken as the error on a given parameter, and the peak of the distribution was taken as the measurement.

The disk fractional luminosity $f_{\mathrm{disk}}=L_{\mathrm{IR}}/L_*$ was calculated from the best-fit blackbody curve and stellar atmosphere model. This quantity can be used to determine the expected disk type (debris, transition or primordial disk; see Section \ref{sec:Discuss}). 

\subsection{Tests and Validations}
\label{subsec:Valid}

%\documentclass[apj]{emulateapj}
%
%\usepackage[latin1]{inputenc}
%%\usepackage[frenchb]{babel}
%\usepackage{graphicx}
%%\usepackage{multicol}
%\usepackage{natbib}
%\usepackage{apjfonts}
%\usepackage{lscape}
%%\usepackage[french, english]{babel}
%%\usepackage{float}
%\usepackage{blindtext}
%\usepackage{url}
%\usepackage{amsmath}
%\usepackage{epstopdf}
%
%\usepackage{tablefootnote}
%\usepackage{threeparttable}
%
%\begin{document}
%%%%%%%%%%%%%%%%%%%%%%%%%%%%%%%%%%%%%%%%%%%%%%%%%%%%%%%%%%%
%
%\clearpage
%%\onecolumngrid
%%\LongTables
%%\begin{landscape}

%\item[•] \textbf{Notes} - $^{\textrm{a}}$ Spectral types taken from \cite{Schneider2012a,Schneider2012b}
%\item[•]  $^{\textrm{b}}$ $T_{\mathrm{eff}}$ estimated from the $T_{\mathrm{eff}}$-Spectral type relation from \cite{Stephens2009} and \cite{Pecaut2013}
%\item[•]  $^{\textrm{c}}$ SEDs fits with models from the BT-Settl grid of \cite{Allard2013}.
%\item[•]  $^{\textrm{d}}$ Does the target have an excess in $W3/W4$ band, according to this work, yes (Y) or no (N)?

%%%%%%%%%%%%%%%%%%%%%%%%%%%
\onecolumngrid
\LongTables
\tabletypesize{\scriptsize}
\begin{deluxetable}{llccccccccc}
\tablewidth{0pt}
\tablecolumns{12}
\tablecaption{TWA members with an excess, reported in \cite{Schneider2012a,Schneider2012b} \label{tab:TWA}}
\tablehead{
\colhead{Name} & \colhead{Spectral Type$^{\rm a}$} & \colhead{$T_{\mathrm{eff, fit}}$ $^{\rm b}$} & \colhead{$S_{W3/W4}$} & \colhead{$p_{W3/W4}$} & \colhead{$W3/W4$} & \colhead{$T_{\mathrm{disk}}$} & \colhead{$f_{\mathrm{disk}}$}  \\
\colhead{ } & \colhead{ } & \colhead{(K)}  & \colhead{($\sigma$)} & \colhead{ } & \colhead{ excess? $^{\rm c}$ }  &  \colhead{(K)} & \colhead{($L_{\mathrm{IR}/L_*}$)}
}
\startdata
   TWA 1  &    K8\,IVe         &  $  4000 \pm  50  $  &  $  44.2 / 38.5  $  &  $   9.7 /  184.6 $  &  Y / Y  &  $  167               \pm 5   $  &  $   0.102   ^{+0.006}_{ -0.003}  $  \\
 TWA 3~AB  &    M4\,IVe         &  $  3000 \pm  60  $  &  $  39.2 / 39.6  $  &  $   6.7 /   51.8 $  &  Y / Y  &  $  216               \pm 5   $  &  $   0.090   ^{+0.003}_{ -0.001}  $  \\
 TWA 4~AB  & K4+K6\,IVe         &  $  3900 \pm  50  $  &  $  47.3 / 43.1  $  &  $  11.2 /  191.2 $  &  Y / Y  &  $  173               \pm 5   $  &  $   0.143   ^{+0.011}_{ -0.005}  $  \\
   TWA 7  &    M3\,IVe         &  $  3300 \pm 180  $  &  $  -3.5 /  3.8  $  &  $   0.8 /    1.5 $  &  N / N  &  $   55  ^{ +130 }_{ - 45 }   $  &  $  \ldots                        $  \\
 TWA 11~B$^{\rm d}$ &     M2\,Ve         &  $  3600 \pm 170  $  &  $  56.2 / 46.0  $  &  $  15.7 /  455.0 $  &  Y / Y  &  $  148               \pm 5   $  &  $   0.45    ^{+0.02 }_{ -0.03 }  $  \\ 
  TWA 27~A &     M8\,Ve         &  $  2400 \pm  50  $  &  $  20.1 /  7.5  $  &  $   3.7 /   14.9 $  &  Y / Y  &  $  300  ^{ +460 }_{ - 40 }   $  &  $   0.064   ^{+0.049}_{ -0.002}  $  \\
  TWA 28  &   M8.5$\,\gamma$ &  $  2000 \pm 130  $  &  $  10.5 /  5.3  $  &  $   2.6 /    9.5 $  &  Y / Y  &  $  290  ^{ +540 }_{ - 50 }   $  &  $   0.128   ^{+0.066}_{ -0.001}  $  \\
 TWA 30~A$^{\rm e}$  &     M5\,Ve         &  $  3000 \pm  70  $  &  $  16.7 / 27.6  $  &  $   2.2 /   13.9 $  &  Y / Y  &  $  203               \pm 8   $  &  $   0.022   \pm 0.001            $  \\
 TWA 30~B$^{\rm e}$  &       M4         &  $  3300 \pm  50  $  &  $  45.7 / 30.6  $  &  $ 322.9 / 2862.8 $  &  Y / Y  &  $  475               \pm 5   $  &  $   4.12    ^{+0.07 }_{ -0.33 }  $  \\
  TWA 31$^{\rm e}$  &     M4.2         &  $  2800 \pm  50  $  &  $  14.6 /  7.7  $  &  $   3.2 /   23.7 $  &  Y / Y  &  $  200  ^{ + 20 }_{ - 15 }   $  &  $   0.047   ^{+0.004}_{ -0.005}  $  \\
  TWA 32$^{\rm e}$  &     M6.3         &  $  2600 \pm 170  $  &  $  12.6 / 23.1  $  &  $   2.4 /   16.1 $  &  Y / Y  &  $  200              \pm 10   $  &  $   0.036   \pm 0.002            $  \\
  TWA 33$^{\rm f}$  &     M4.7         &  $  2900 \pm  50  $  &  $  18.9 / 22.3  $  &  $   2.7 /   12.6 $  &  Y / Y  &  $  240              \pm 10   $  &  $   0.026   \pm 0.001            $  \\
  TWA 34$^{\rm f}$  &     M4.9         &  $  2700 \pm  50  $  &  $  15.2 / 12.2  $  &  $   2.1 /    8.0 $  &  Y / Y  &  $  250              \pm 20   $  &  $   0.020   ^{+0.001}_{ -0.002}  $  
\enddata

\end{deluxetable}
%
%
%
%\clearpage
%%\end{landscape}
%\twocolumngrid
%
%
%%%%%%%%%%%%%%%%%%%%%%%%%%%%%%%%%%%%%%%%%%%%%%%%%%%%%%%%%%%
%
%\end{document}
\begin{tablenotes}
\footnotesize
\item[•] \textbf{Notes} - $^{\textrm{a}}$ Spectral types are taken from \cite{Schneider2012a,Schneider2012b} and \cite{Pecaut2013}.
\item[•]  $^{\textrm{b}}$ SED fits with models from the BT-Settl grid of \cite{Allard2013}.
\item[•]  $^{\textrm{c}}$ Excess in $W3/W4$ band, as found in this work, yes (Y) or no (N).
\item[•]  $^{\textrm{d}}$ AllWISE photometry was used for this target, as the WISE data seems to correspond to TWA\,11~A instead.
\item[•]  $^{\textrm{e}}$ \cite{Schneider2012a} find disk temperatures of $210$, $190+660$ (two blackbodies were fit), $200$ and $200\,$K for TWA 30\,A, 30\,B, 31 and 32, respectively.
\item[•]  $^{\textrm{f}}$ \cite{Schneider2012b} find disk temperatures of $190+850$ and $210+900\,$K for TWA 33 and 34, respectively. (See Table~\ref{tab:2temp} for comparison with our results)
\item[•]
\end{tablenotes}
\twocolumngrid

%To verify that our algorithm worked well, we checked if we could find the TW Hya stars that are known to have disks within all the members of the association that were compatible with our target sample, i.e. members colder than $5000\,$K, which excludes TWA 11A and 19A. We based our comparison on the results of \cite{Schneider2012a} and \cite{Schneider2012b}. 
The method presented here was validated by analyzing all members of TWA cooler than 5000\,K with known disks. The results were then compared to those of \cite{Schneider2012a} and \cite{Schneider2012b}. We considered all objects that have a $W3$ and $W4$ excess value $>5$ (see the fourth to last and the last columns of their Table~2; \citealt{Schneider2012a}) as displaying a significant excess. Using the method presented in Section~\ref{subsec:InitDetect}, we were able to retrieve all 12 stars as significant excess candidates: TWA~1, 3\,AB, 4\,AB, 11\,B, 27, 28, 30\,A, 30\,B, 31, 32, 33 and 34), except for one, TWA~7. This last object was rejected by our method because its values of $p_{W3}$ \emph{and} $p_{W4}$ were too low, respectively $\sim 0.8$ and 1.5, even though its $W4$ excess level is present at $3.8\,\sigma$. The results of this analysis are listed in Table~\ref{tab:TWA}. TWA~7 is known to have a debris disk with an inner disk at $66\,$K and a colder outer disk at $20\,$K \citep{Riviere2013}, which was recently directly imaged for the first time by \cite{Choquet2015}. This demonstrates that our method may miss fainter/colder disks, but is efficient at identifying brighter/warmer disks. Moreover, we did not flag any other TWA star as a disk candidate, which is consistent with the findings of \cite{Schneider2012a,Schneider2012b}. % In other words, even though we did detect a significant excess for TWA 7, its relative excess level was too small according to our criteria.

\section{Observations and data reduction}
\label{sec:Obs}
New follow-up observations were obtained for two of the new disk candidates identified here. These observations are described in the remainder of this section.

\subsection{Optical spectroscopy with ESPaDOnS}
\label{subsec:espadons}
High-resolution optical spectroscopy was obtained for 2MASS~J05010082--4337102 in queue service observing (QSO) mode with ESPaDOnS \citep{Donati2006} at CFHT on 2016 March 3. ESPaDOnS was used in the Spectro-Polarimetric mode combined with the “slow” CCD readout mode, yielding a resolving power of R $\sim 68000$ in the range 3700--$10500\,\angstrom$. A total integration time of 90 minutes was used. The data were reduced by the QSO team using the CFHT pipeline UPENA1.0. This pipeline uses J-F. Donati’s software Libre-ESpRIT \citep{Donati1997}. 

\subsection{Near-infrared spectroscopy with FIRE}
\label{subsec:fire}
A high-resolution near-infrared spectrum was obtained for 2MASS~J02265658--5327032 with the Folded-port InfraRed Echellette (FIRE; \citealp{2008SPIE.7014E..0US,2013PASP..125..270S}) spectrograph at the Magellan Baade telescope on 2015 September 24. The 0\farcs6 slit was used, yielding a resolving power of $R \sim 6\,000$ over 0.8--2.45\,$\mu$m. Two 650\,s exposures were obtained, resulting in a signal-to-noise ratio of $\sim$\,43.8 per pixel at 1.51--1.55\,$\mu$m. The standard A0-type star HD~25986 was observed immediately before the science target (4 $\times$ 90\,s) to achieve telluric correction. A wavelength calibration ThAr lamp exposure was also obtained immediately before the science target, and standard flat field calibrations were obtained at the beginning of the night. The data were reduced using the Firehose~v2.0 package \citealp{2009PASP..121.1409B,zenodofirehose}\footnote{Available at \url{https://github.com/jgagneastro/FireHose\_v2/tree/v2.0}}; see \citealt{BANYANVII} for more detail on this reduction package. The Firehose package automatically corrects the wavelength solution to vacuum wavelengths in a heliocentric frame of reference.

\section{Results}
\label{sec:Results}

\subsection{Candidates with excess}
\label{subsec:Candidate}
The analysis described above was applied on the bona fide and candidates targets of \cite{Malo2013} and \cite{Malo2014b}, which mainly contain earlier-type stars (K5V--M5V) than those of \cite{BANYANVII}. From the total sample of 369 (with 261 objects following the quality criteria), only 6 went through the inital excess detection phase. One of them was contaminated in every WISE images. The other 5 displayed convincing excesses, but were already known in the literature : (1) Fk Ser, a binary T Tauri star in the Argus association with a known T Tauri disk \citep{Jensen1996}, (2) TWA 3 and (3) TWA 30~A, (4) DZ Cha, a member of the $\sim 6\,$Myr old $\epsilon$ Cha association \citep{Torres2008} with a known debris disk \citep{Wahhaj2010,Simon2012}, and (5) V4046 Sgr, a pre-main-sequence spectroscopic binary (T Tauri star) with a circumbinary disk \citep{Jensen1996,Jensen1997}. No new disk candidates were thus found in the \cite{Malo2013,Malo2014b} sample.

\onecolumngrid
\LongTables
\tabletypesize{\scriptsize}
\begin{deluxetable}{lccccc}
\tablewidth{0pt}
\tablecolumns{6}
\tablecaption{Candidate Properties \label{tab:top6_01}}
\tablehead{
\colhead{2MASS designation} & \colhead{J05010082--4337102} & \colhead{J08561384--1342242} & \colhead{J12474428--3816464} & \colhead{J02265658--5327032} & \colhead{Ref.}
}
\startdata
RA (deg) & $       75.253587$ & $       134.05750$ & $       191.93441$ & $       36.736258$ & $1$ \\
DEC (deg) & $      -43.619529$ & $      -13.706796$ & $      -38.279607$ & $      -53.450924$ & $1$ \\
Spectral Type & M4.5 & M8$\,\gamma$ & M9$\,\gamma$ & L0$\,\delta$ & $2$ \\
Association & COL & TWA & TWA & THA & $2$ \\
Age (Myr) & $42^{+6}_{-4}$ & $10\pm 3$ & $10\pm 3$ & $45\pm 4$ & $3$ \\
$I$ (mag)   & $   13.4 \pm     0.03$  & $   \ldots$ 		    & $   17.9   \pm    0.2$  & $   \ldots$ & $4$ \\
$J$ (mag)   & $   11.61 \pm     0.03$ & $   13.60 \pm     0.03$ & $   14.78 \pm     0.03$ & $   15.40 \pm     0.05$ & $1$ \\
$H$ (mag)   & $   11.06 \pm     0.02$ & $   12.98 \pm     0.03$ & $   14.10 \pm     0.04$ & $   14.35 \pm     0.05$ & $1$ \\
$K_s$ (mag) & $   10.75 \pm     0.02$ & $   12.49 \pm     0.02$ & $   13.57 \pm     0.04$ & $   13.75 \pm     0.05$ & $1$ \\
$W1$ (mag)  & $   10.54 \pm     0.02$ & $   12.15 \pm     0.02$ & $   13.11 \pm     0.02$ & $   13.22 \pm     0.02$ & $5$ \\
$W2$ (mag)  & $   10.29 \pm     0.02$ & $   11.62 \pm     0.02$ & $   12.52 \pm     0.02$ & $   12.78 \pm     0.03$ & $5$ \\
$W3$ (mag)  & $    9.32 \pm     0.02$ & $    9.88 \pm     0.05$ & $   10.95 \pm     0.08$ & $   11.6 \pm     0.1$ & $5$ \\
$W4$ (mag)  & $    7.12 \pm     0.07$ & $    8.4 \pm      0.3$  & $    8.8 \pm      0.3$  & $    8.6 \pm     0.3$ & $5$ \\
$W1-W4$ (mag) & $  3.42 \pm  	0.08$ & $ 	 3.8 \pm      0.3$  & $    4.3 \pm      0.3$  & $    4.6 \pm     0.3$ & $9$ \\
Stat. Distance (pc) & $   47.8^{+    7.2}_{-    8.4}$ & $   35.8 \pm     4.4$ & $   76.6^{+    8.4}_{-    8.0}$ & $   43.4^{+    2.8}_{-    2.4}$ & $2$ \\
Estim. $M_*$ ($M_{\mathrm{Jup}}$) $^{\rm a}$& $  119^{+   19}_{-   17}$ & $   14.4^{+    0.8}_{-    1.4}$ & $   17.4^{+    0.8}_{-    0.9}$ & $   13.7 \pm     0.3$ & $2$ \\
$T_{\mathrm{SpType}}$ (K) & $    3125$ & $    2550$ & $    2400$ & $    2260$ & $6,7$ $^{\rm b}$ \\
$T_{\text{BANYAN VII}}$ (K) $^{\rm c}$& $  \ldots  $ & $    2300 \pm      200$ & $    2100 \pm      300$ & $    1700 \pm      100$ & $8$ \\
$T_{\mathrm{eff, fit}}$ (K) $^{\rm d}$& $    2900 \pm 50$ & $    2100 \pm 50$ & $    2000 \pm 50$ & $    1800 \pm 50$ & $9$ \\
$\log{g_{\mathrm{ fit}}}$ $^{\rm d}$& $4.00 \pm 0.60$ & $4.00 \pm 0.30 $ & $5.00 \pm 0.60 $ & $4.50 \pm 0.60$ & $9$ \\
$T_{\mathrm{disk}}$ (K) & $     170 \pm 10$ & $     520^{+25}_{-275} $ & $     190^{+55}_{-10} $ & $     135 \pm 20$ & $9$ \\
%$T_{\mathrm{disk}}$ (K) & $     173^{+3}_{-9}$ & $     530^{+18}_{-287}$ & $     185^{+307}_{-5} $ & $     134^{+11}_{-21 }$ & $8$ \\
$f_{\mathrm{disk}}$ & $0.021 ^{+0.002}_{-0.001} $ & $0.089  ^{+0.011}_{-0.016} $ & $0.087  ^{+0.023}_{-0.010} $ & $0.15  ^{+0.06}_{-0.04} $ & $9$ \\
$S_{W3/W4}$ ($\sigma$) & $ 9.7 / 13.5$ & $14.1 /  3.7$ & $ 8.0 /  3.5$ & $ 2.2 /  4.0$ & $9$ \\
$p_{W3/W4}$ & $ 1.7 / 14.1 $ & $ 3.6 / 15.0 $ & $ 2.8 / 20.3 $ & $ 1.5 / 22.6 $ &  $9$ 
%$p_{W3/W4}$ & $ 1.611 / 12.938 $ & $3.215 / 13.195$ & $2.950 / 21.846$ & $ 1.522 / 22.840 $ &  $8$ \\
\enddata

\end{deluxetable}

\begin{tablenotes}
\footnotesize
\item[•] \textbf{Notes} - $^{\textrm{a}}$ The estimated masses were calculated by comparing the 2MASS and WISE photometry with the AMES-Cond isochrones from \cite{Baraffe2003}, assuming the YMG age and the trigonomic or statistical distance of BANYAN~II \citep{BANYANII}.
\item[•]  $^{\textrm{b}}$ $T_{\mathrm{eff}}$ estimated from the $T_{\mathrm{eff}}$--Spectral type relation from \cite{Stephens2009} and \cite{Pecaut2013}.
\item[•]  $^{\textrm{c}}$ $T_{\mathrm{eff}}$ estimated by \cite{BANYANVII}.
\item[•]  $^{\textrm{d}}$ Models from the BT-Settl grid of Allard et al. (2013). 
\item[•] \textbf{References} - (1) 2MASS catalog, \citealt{2MASS}; (2) BASS catalog \citealt{BANYANV}; (3) \citealt{Bell2015}; (4) DENIS catalog, \citealt{DENIS}; (5) WISE All-Sky Source catalog, \citealt{WISE}; (6) \citealt{Stephens2009}; (7) \citealt{Pecaut2013}; (8) \citealt{BANYANVII}; (9) this work. %**BANYAN~II
\item[•]
\end{tablenotes}
\twocolumngrid

The same analysis was performed on the initial 1228 objects from \cite{BANYANV}, with only 324 satisfying the quality requirements. This number was reduced to 53 after the first initial excess detection, then to 35 after a visual inspection of the $W3$ channel images, and then to only 14 after a visual inspection of the $W4$ channel images. After inspecting the 2MASS images, 4 other targets were eliminated, leaving a total of 10 circumstellar disk candidates. Among these objects, 6 were already known to have a disk in the literature: 2MASS J02590146--4232204 \citep{Rodriguez2013}, TWA 27, TWA 28, TWA 32 \citep{Schneider2012a} as well as TWA 33 and TWA 34 \citep{Schneider2012b}. These targets were thus set aside for the rest of this work. The final sample contains 4 new circumstellar disk candidates; their properties are listed in Table~\ref{tab:top6_01}. It should be noted that the excess of J1247--3816 was previously identified in \cite{Rodriguez2015}, but had not been explicitly characterized by them.

The four new disk candidates identified in this work have mid-M to early-L type host stars, which are also candidates of YMGs. Two of them (J0856--1342 and J1247--3816) are candidate members of TWA ($10\pm 3\,$Myr). It is not surprising to find TWA candidates with IR-excess and circumstellar disks since almost half of the stars in this association display an IR excess. In fact, the 22 $\mu$m excess fraction of this young association has been revised at $42^{+10}_{-9}\%$  by \cite{Schneider2012a} from the previous work of \cite{Looper2010}, \cite{Rebull2008} and \cite{Low2005}. The other two disk candidates (J0501--4337 and J0226--5327) are candidate members of COL ($42^{+6}_{-4}\,$Myr) and THA ($45\pm 4\,$Myr), respectively. %We are also not surprised that we didn't find disks in older moving groups (like ABDor; $110-130\,$Myr), since the disks lifetime is much shorter ($<40\,$Myr; see Section \ref{subsec:disknature}).

The Canadian Astronomy Data Center (CADC; which includes the Hubble Space Telescope (HST), the Gemini Observatory and the Canada-France-Hawaii Telescope (CFHT) data) were parsed for additional information on the four new disk candidates presented here. A similar search was performed in the ESO, Keck and Subaru astronomical archives. Only SOFI images for J0856--1342 and J1247--3816 were found. We verified whether contamination could be detected around the targets within an aperture of 16.5$''$: no such contamination was found. Additional, ALMA (\textit{Atacama Large Millimeter/submillimeter Array}) data were found for J1247--3816 (with the ALMA band number 6, at 211--275\,GHz; corresponding to 1.42--1.09 mm). The results from these observations were analyzed and presented in \cite{Rodriguez2015}. There was no detection at the position of our target at these wavelengths, with a corresponding continuum flux upper limit of 0.15\,mJy, suggesting an absence of small cold dust grains in this system.%(except for a continuum source that was offset by 10’’ from our target, which is probably a background source)

\begin{figure}
  \begin{tabular}{@{}cc@{}}
    \includegraphics[scale=0.41]{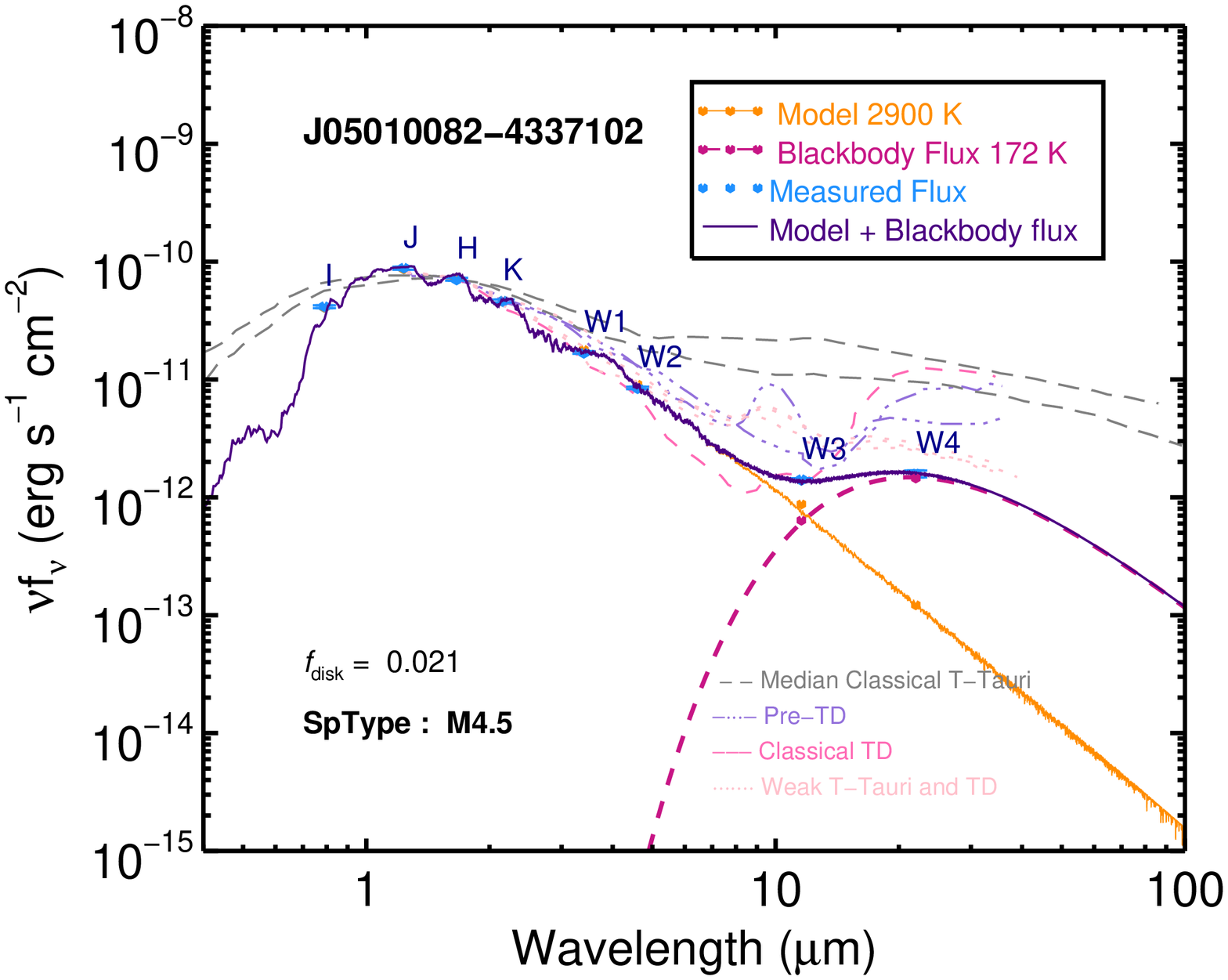} &  % scale=0.41
    \includegraphics[scale=0.41]{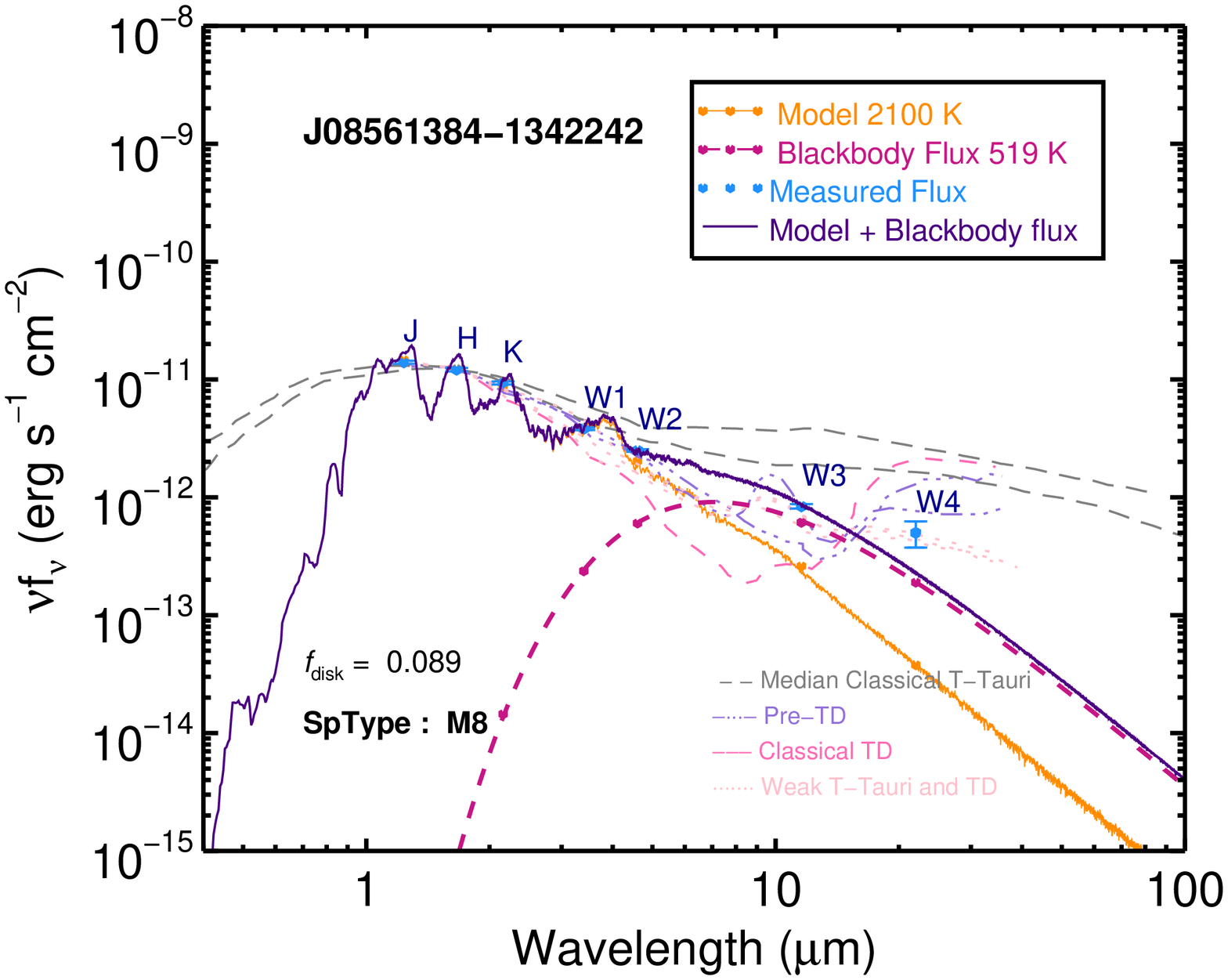} \\
    \includegraphics[scale=0.41]{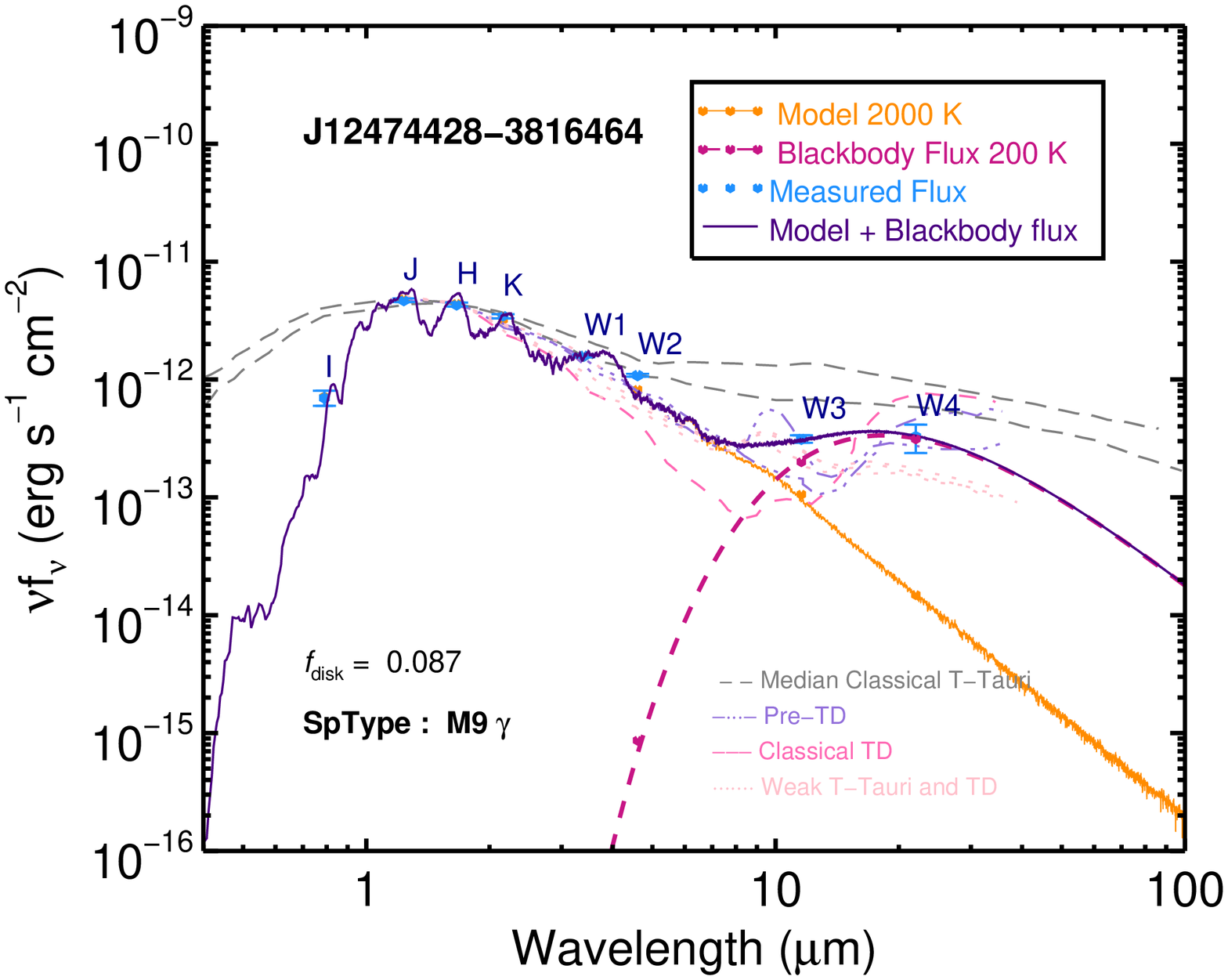} &
    \includegraphics[scale=0.41]{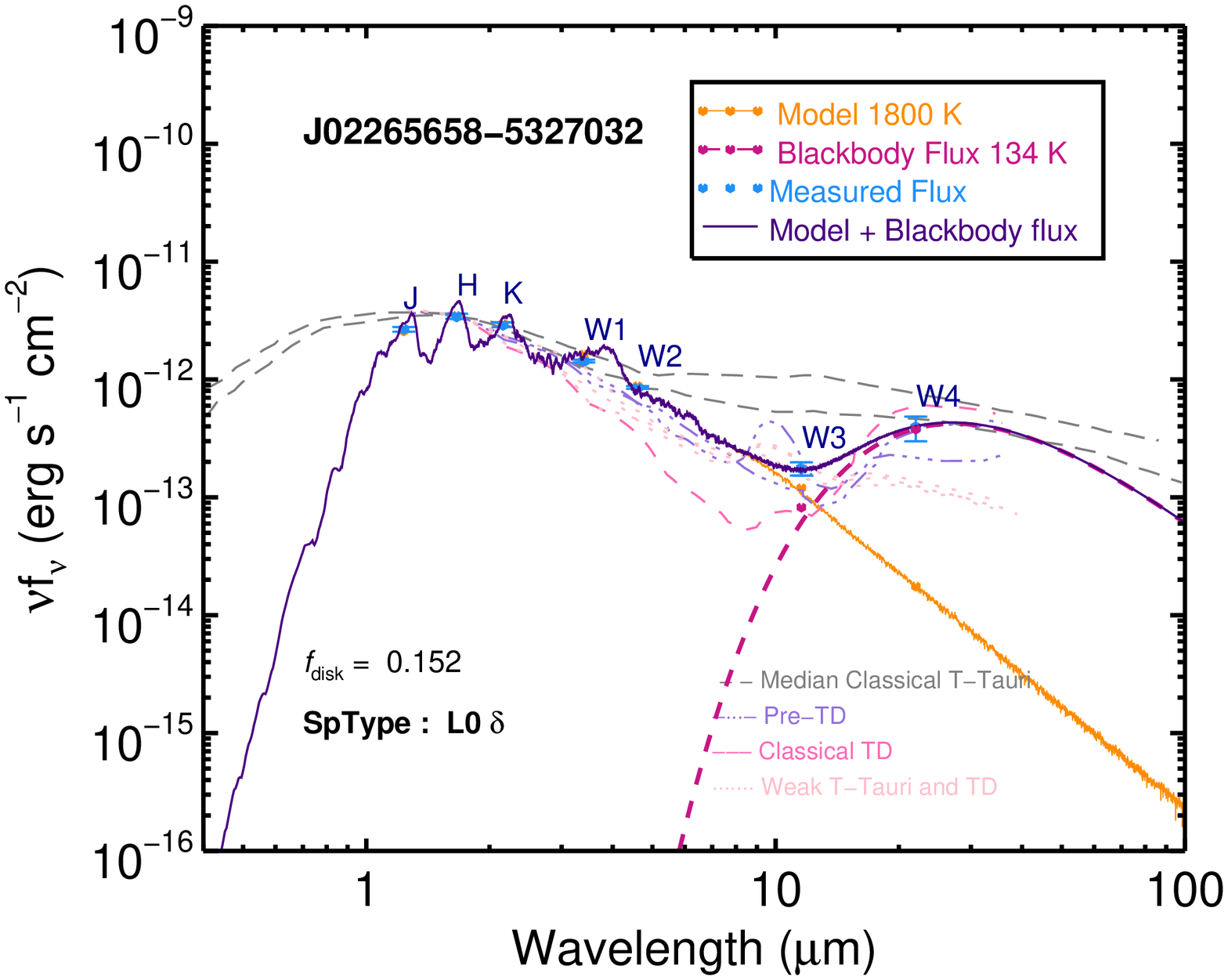} 
  \end{tabular}
%\centering

\caption{Observed and modeled spectral energy distributions (SEDs) of J0501--4337, J0856--1342, J1247--3816 and J0226--5327. The SED of different types of disks have been added to see how these systems compare in properties with one another. The classical TD is from the CoKu Tau/4 star (pink), the pre-TDs are UX Tau and LkCa 15 (lilac), and the weak-TD is FQ Tau (light pink); all SEDs were taken from \cite{Najita2007}. The upper and lower quartiles of the median T Tauri SEDs (grey) are from \cite{Najita2007}, which are originally from \cite{Dalessio1999}. The median SEDs for weak-T Tauri in Taurus and Chamaeleon I (light pink) were taken from a presentation given by Elise Furlan, for the 2011 Hubble Fellows Symposium\footnote{\url{https://webcast.stsci.edu/webcast/detail.xhtml;jsessionid=201F62F7409A2DA45E9387E3E524CC72?talkid=2482&parent=1}}. The model spectra correspond to those yielding the lowest $\chi^2_{\mathrm{red}}$ values over the ($I$), $J$, $H$ \& $K_s$ bands. The orange curve represents the flux density from the synthetic spectrum, the magenta curve is the flux density from the blackbody spectrum and the indigo one is the total flux. Their respective $T_{\mathrm{eff}}$ are given in the legend. Each point on these curves represents the synthetic flux associated with each photometric band pass for comparison with the observed photometry (blue points). \\
% to better compare them to the observed one, which are shown in blue.
 }
 \label{fig:BBfit}
\end{figure}

In Figure~\ref{fig:BBfit}, we display the observed and modeled spectral energy distributions (SEDs) of our 4 candidates from their best single-backbody fit. Representative SEDs from classical and weak T Tauri, pre-transition, classical and weak TD have been added for comparison. The comparison SEDs have been scaled to the $H$ band flux density. All four disk candidates have SEDs that differ significantly from those of T Tauri-type objects. The SED of J0501--4337 has a behavior similar to that of a classical TD, but with a lower 22$\,\mu$m flux. The fact that this target does not display a T Tauri SED is consistent with the fact that its H$\alpha$ EW is weaker than the T Tauri criterion of \cite{Barrado2003}. The SED of J0856--1342 is similar to those of pre-transition or weak-T Tauri disks, while that of J1247--3816 definitely seems like a pre-transition disk. The SED of J0226--5327 could be caused by either a classical or pre-transition disk. The addition of data points between $W2$ and $W3$ as well as redwards of $W4$ would be helpful to further determine the nature of these disks.

The SEDs of J0501--4337 and J0226--5327 are well reproduced across the whole spectral range, while those of J0856--1342 and J1247--3816 do not simultaneously agree with the $W2$ and $W4$ data points. The same behavior has been observed with the validation performed on the TWA members.  

%\documentclass[apj]{emulateapj}
%
%\usepackage[latin1]{inputenc}
%%\usepackage[frenchb]{babel}
%\usepackage{graphicx}
%%\usepackage{multicol}
%\usepackage{natbib}
%\usepackage{apjfonts}
%\usepackage{lscape}
%%\usepackage[french, english]{babel}
%%\usepackage{float}
%\usepackage{blindtext}
%\usepackage{url}
%\usepackage{amsmath}
%\usepackage{epstopdf}
%
%\usepackage{tablefootnote}
%\usepackage{threeparttable}
%
%\begin{document}
%%%%%%%%%%%%%%%%%%%%%%%%%%%%%%%%%%%%%%%%%%%%%%%%%%%%%%%%%%%
%
%\clearpage
%%\onecolumngrid
%%\LongTables
%%\begin{landscape}

%\item[?] \textbf{Notes} - $^{\textrm{a}}$ Spectral types taken from \cite{Schneider2012a,Schneider2012b}
%\item[?]  $^{\textrm{b}}$ $T_{\mathrm{eff}}$ estimated from the $T_{\mathrm{eff}}$-Spectral type relation from \cite{Stephens2009} and \cite{Pecaut2013}
%\item[?]  $^{\textrm{c}}$ SEDs fits with models from the BT-Settl grid of \cite{Allard2013}.
%\item[?]  $^{\textrm{d}}$ Does the target have an excess in $W3/W4$ band, according to this work, yes (Y) or no (N)?

%%%%%%%%%%%%%%%%%%%%%%%%%%%
\onecolumngrid
\LongTables
\tabletypesize{\scriptsize}
\begin{deluxetable}{lcccccccc}
\tablewidth{0pt}
\tablecolumns{8}
\tablecaption{Two-temperature fits properties\label{tab:2temp}}
\tablehead{
\colhead{Name} &  \colhead{$S_{W2}$} & \colhead{$p_{W2}$}  & \colhead{$T_{\mathrm{disk}}$} &   \colhead{$T_{\text{disk 1}}$} & \colhead{$T_{\text{disk 2}}$ $^{\text{ a}}$}  & \colhead{$f_{\mathrm{disk 1 + disk 2}}$} &  \colhead{$T_{\text{disk 1}}$} & \colhead{$T_{\text{disk 2}}$ $^{\text{ b}}$}  \\
\colhead{} & \colhead{($\sigma$)} & \colhead{ } & \colhead{(K)} &\multicolumn{2}{c}{(this work)} & \colhead{ } & \multicolumn{2}{c}{\citep{Schneider2012b}} 
}
\startdata
  J08561384--1342242 & $ 4.7 $ & $ 1.3 $ & $ 530 ^{ + 20}_{ -290} $ & $ 510 - 700 $ & $  50 - 220 $ &  $ 0.11  - 1.40 $ & $ \ldots $ & $ \ldots $ \\
  J12474428--3816464 & $ 4.4 $ & $ 1.3 $ & $ 185 ^{ +310}_{ -  5} $ & $ 510 - 600 $ & $  50 -  60 $ &  $ 0.72  - 5.65 $ & $ \ldots $ & $ \ldots $ \\
              TWA 33 & $ 3.2 $ & $ 1.2 $ & $ 240 ^{ + 5}_{ -  10} $ & $ 580 - 940 $ & $  70 - 210 $ &  $ 0.047 - 0.16 $ & $    850 $ & $    190 $ \\
              TWA 34 & $ 6.2 $ & $ 1.5 $ & $ 250 ^{ + 10}_{ - 15} $ & $ 840 - 930 $ & $  50 -  80 $ &  $ 0.11  - 0.46 $ & $    900 $ & $    210 $ 
\enddata

\end{deluxetable}
%
%
%
%\clearpage
%%\end{landscape}
%\twocolumngrid
%
%
%%%%%%%%%%%%%%%%%%%%%%%%%%%%%%%%%%%%%%%%%%%%%%%%%%%%%%%%%%%
%
%\end{document}
\begin{tablenotes}
\footnotesize
\item[•] \textbf{Notes} - $^{\text{a}}$ Disk temperatures obtained from a two-temperature fit, from this work.
\item[•]  $^{\text{b}}$ Disk temperatures obtained from a two-temperature fit from \cite{Schneider2012b}.
\item[•]
\end{tablenotes}
\twocolumngrid

\onecolumngrid

\twocolumngrid
 
This behavior may indicate that there is both a hot and a cold disk component, requiring a fit with two blackbodies at different temperatures; this was thus performed on these objects. The results from these fits are listed in Table~\ref{tab:2temp}, as well as the results from \cite{Schneider2012b} for comparison and validation of our method. 
 
An alternative explanation for the poor fit of the $W2$/$W4$ data points for J0856--1342 and J1247--3816 could be that the predicted spectral model fluxes at $W2$ are less accurate in this age and temperature range (both are $\sim 10\,$Myr and $\sim2000\,$K). For this reason, the excess were also fitted using only $W3$ and $W4$. Mean disk temperatures of $260^{+60}_{-10}$ and $200^{+40}_{-5}\,$K were obtained for J0856--1342 and J1247--3816, respectively. 

\subsection{New spectroscopy of candidates}
A heliocentric radial velocity of $19.6\pm 0.5\,$km\,s$^{-1}$ and a projected rotational velocity of $11\,$km\,s$^{-1}$ were measured from the ESPaDOnS spectrum of J0501--4337, using the method described by \cite{Malo2014a}. J0501--4337 displays lithium absorption with equivalent width (EW) of EW$_{\rm Li} = 230\pm 30\,$m$\angstrom$, measured using the method described by \cite{Malo2014b}. The spectrum also displays strong H$\alpha$ emission, with EW$_{\rm H\alpha} = 7.56 \pm 0.20\,\angstrom$.

The full width at 10\% intensity of the H$\alpha$ emission line (H$\alpha_{10\%}$) can be used as a tracer of accretion \citep{Natta2004}; it was measured on two separate orders of the ESPaDOnS spectrum. A running smoothing filter of 10 pixels was applied first to mitigate the effect of bad pixels, and H$\alpha_{10\%}$ was calculated in a $10^4$ steps of a Monte Carlo simulation where the spectral flux density varied according to its measurement errors. The two-parameters relation of \cite{Natta2004} was used to calculate the corresponding accretion rate at each Monte Carlo step. Since \cite{Natta2004} provide error bars on their H$\alpha_{10\%}$--accretion rate relation parameters, they were also made to vary according to their error bars. Individual probability density functions (PDFs) were created for both measurements in each order. Final measurement PDFs were obtained from a multiplication of the per-order PDFs.

This analysis has yielded H$\alpha_{10\%} = 210.7^{+7.6}_{-2.2}\,$km\,s$^{-1}$ and a corresponding accretion rate of:
\begin{align}
\log _{10} \dot{M}_{\mathrm{ac}} = -10.80^{+0.07}_{-0.05}\,M_\odot\,\mathrm{yr}^{-1}.
\end{align}

\cite{Natta2004} use a threshold of H$\alpha_{10\%} \geq 200\,$km\,s$^{-1}$ to determine whether a substellar or low-mass object is accreting; 99.3\% of the Monte Carlo steps described above respected this criterion. In the present case, it cannot be ruled out with certainty that the H$\alpha$ emission is due to chromospheric activity rather than accretion.

A radial velocity measurement was obtained from the FIRE spectrum of J0226--5327, using a forward modeling method based on the zero-velocity, solar metallicity CIFIST2011 BT-Settl models \citep{Baraffe2003,2012RSPTA.370.2765A}. The best-fitting model was selected by minimizing $\chi^2$ over several high-S/N orders, and was then Doppler-shifted (1 free parameter), convolved with a Gaussian instrumental line spread function (1 free parameter for the Gaussian width) and multiplied with a linear slope (2 free parameters) to account for instrumental effects. The Interactive Data Language (IDL) implementation of the amoeba Nelder-Mead downhill simplex algorithm \citep{Nelder1965in} was used to minimize the $\chi^2$ difference between the forward model and the data in fifteen 0.02\,$\mu$m-wide windows regularly distributed between 1.5100--1.5535\,$\mu$m in the $H$ band, where there is a large S/N and a good radial velocity information content. This method is based on A.~J. Burgasser et al. (in preparation; see also \citealt{2015ApJ...808L..20G,Kellogg2016uq}) and ensures that the results are robust against bad pixels and other instrumental systematics. The median and standard deviations of these 15 measurements yielded a radial velocity of $4.7 \pm 1.4$\,km\,s$^{-1}$, to which a $\pm 3$\,km\,s$^{-1}$ systematic error (calculated from comparison with radial velocity standards; \citealt{Kellogg2016uq}) is added in quadrature to obtain a final radial velocity measurement of $4.7 \pm 3.3$\,km\,s$^{-1}$. Moreover, the spectrum of J0226--5327 displays clear signs of Paschen $\beta$ emission (Figure~\ref{fig:PaBeta}).

\section{Discussion}
\label{sec:Discuss}

\subsection{Evidence for youth and membership in young associations}
\label{subsec:youth}

\begin{figure}
\centering
  \begin{tabular}{@{}cc@{}}
    \includegraphics[scale=0.55]{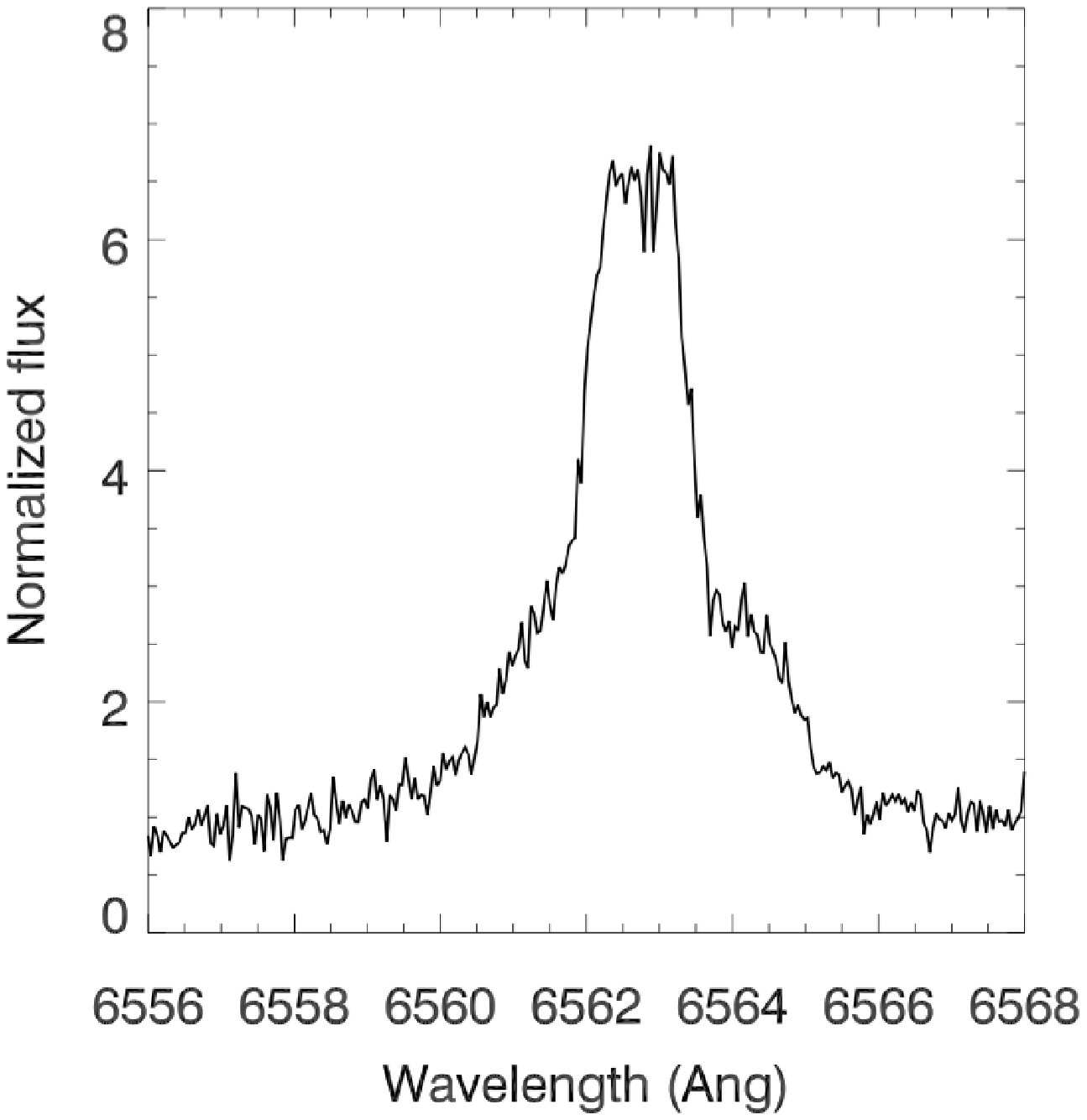} \\
    \includegraphics[scale=0.55]{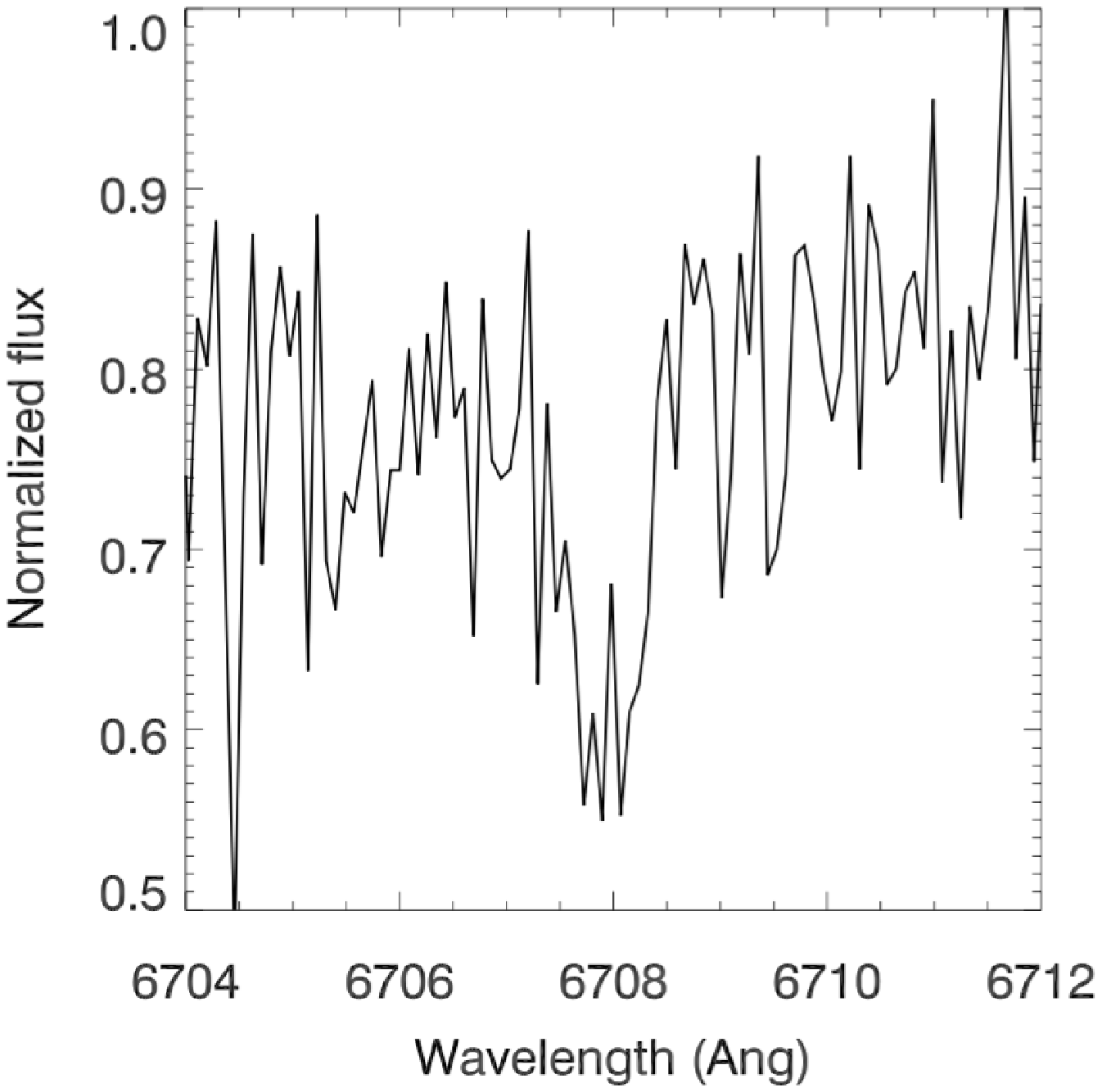} 
  \end{tabular}
\caption{Top panel: H$\alpha$ emission line of J05010082--4337102. Bottom panel: lithium absorption line for J05010082--4337102. Both are consistent with a young age. See Section \ref{subsec:youth} for more details.}
 \label{fig:J0501_lines}
\end{figure}

\textbf{2MASS~J05010082--4337102: } Using the newly obtained radial velocity measurement of $19.6\pm 0.5\,$km\,s$^{-1}$ in BANYAN~II, this M4.5 star is a candidate member of COL (59\%), THA (10\%) or $\beta$PMG (28\%). The first two YMGs have similar ages ($42^{+6}_{-4}\,$Myr for COL and $45\pm4\,$Myr for THA) while $\beta$PMG is slightly younger ($24\pm 3\,$Myr; \citealt{Bell2015}). The presence of lithium absorption in its optical spectrum\footnote{Lithium depletion boundary available at \href{https://figshare.com/articles/Lithium_depletion_boundary_for_low_mass_stars_and_brown_dwarfs/1326730}{this link}. Figure from J.~Gagné, based on \cite{Basri1998} and the models of \cite{Baraffe2003}.} is a sign of youth (see Figure~\ref{fig:J0501_lines}). \cite{Kraus2014} determine the location of the LDB at SpT=M$4.5\pm0.3$ for THA ($45\pm4\,$Myr), while \cite{Binks2014} put the LDB at SpT=M$4.475\pm0.525$ for $\beta$PMG ($24\pm3\,$Myr). At $232 \pm 28\,$m$\angstrom$, the Li EW of J0501--4337 is consistent with a partial depletion (undepleted sources at this SpT typically have $>500\,$m$\angstrom$), and given its M4.5 spectral type it is thus compatible with being on/near the LDB of THA or BPMG per the above results. Indeed, the Li EW of this source is in excellent agreement with those of other M4.5 members of THA. There are currently only two BPMG members at M4.5, and both have an EW of $\sim500\,$m$\angstrom$, higher than for J0501--4337; this could indicate an older age for the latter, but the statistical significance is low given the small numbers. According to \cite{Barrado2004}, the LDB for the slightly older association IC~2391 ($\sim 50\,$Myr) is at $M_{I_c}\simeq10.2\,$mag, while for our target $M_{I_c}=10.03\,$mag. The presence of Li in J0501--4337 may thus indicate that it is slightly younger than 50\,Myr. %and H$\alpha$ emission (if from accretion as suspected) are both 

\textbf{2MASS~J08561384--1342242: } This M8$\,\gamma$ was proposed as a low-probability candidate member of TWA with a 4.9\% membership probability and <0.1\% Field Contamination Probability (FCP). Even though the Bayesian membership probability is low, there are clear indications that this object is young, it has a very low surface gravity ($\gamma$ suffix). According to \cite{Kirkpatrick2006}, \cite{Cruz2009} and \cite{Allers2013}, all members of associations younger than 200\,Myr display signs of low gravity, while older objects do not. Given the youth of this star and its low probability in TWA, it is possible that it is a member of another young association that is not yet included in the BANYAN~II tool.%, but a photometric distance is needed to confirm this hypothesis.

\textbf{2MASS~J12474428--3816464: } This M9$\,\gamma$ brown dwarf has been proposed as a candidate member of TWA (46.6\%) with <0.1\% FCP \citep{Gagne2014b}. This target was shown to have a lower-than-normal equivalent width of atomic species in its visible and NIR spectra, a triangular-shaped $H$-band continuum, and redder-than-normal NIR colors for its spectral type, all of which indicate that it is young \citep{Allers2013,Cruz2009}. 

There are some discrepancies in the kinematics of this star from that of TWA, explaining its modest membership probability. This brown dwarf could also be a member of a young association not yet included in the spatial and kinematics models from BANYAN~II, such as the Lower Centaurus Crux (LCC) association (Gagn\'e et al. 2016, \textit{submitted}). LCC members have similar UVW kinematics to those of TWA, but are a little further away from the Sun (with distances of $\sim 85-150\,$pc for LCC, compared to $\sim35-70\,$pc for TWA). 

It has been suggested that TWA and LCC could be two parts of the same group  \citep{Song2003,Zuckerman2004,Schneider2012a,Schneider2012b}. While the kinematics of this target are a good match to LCC, if placed at such a large distance it would be abnormally bright even for a young dwarf. A trigonometric distance measurement will be needed to resolve this. 

\textbf{2MASS~J02265658--5327032: } Including the new radial velocity measurement to previous observables in the BANYAN~II tool yields a probability of 99.96\% that J0226--5327 is a member of the Tucana-Horologium association (THA) with a most probable statistical distance of $43.4^{+2.8}_{-2.4}$\,pc (assuming THA membership). Only a trigonometric distance measurement is needed to confirm this target as a bona fide member of THA. This very late-type object (L0$\,\delta$) is also one of the very low-surface gravity ($\delta$) objects from the BASS survey \citep{Faherty2016}. %: its H-band has a triangular shape, both of which are youth indicators. Plus, like the previous target, it has redder than normal colors. %***Peculiar spectrum (selon Jo)? 

The presence of Pa$\beta$ emission is a sign of ongoing accretion, which means that there is still gas in the disk, and further indicates that this brown dwarf is young. This target has a $W4$ upper limit flag in the WISE \emph{aperture photometry} data set, but not in the \emph{profile-fit} one. 

In addition to the above considerations, the presence of excess infrared emission in these four objects provides further evidence for their youth : \cite{Zuckerman2004} and \cite{Moor2006} argue that stars with $f_{\mathrm{disk}} > 10^{-3}$  are younger than $100\,$Myr.

\subsection{Excess and disk characteristics}
\label{subsec:disknature}
Table~\ref{tab:top6_01} reports the value of the fractional luminosity $f_{\mathrm{disk}}$ for each target; they range between 0.021 and 0.151. Objects with $f_{\mathrm{disk}} \lesssim 0.01$ are typically considered debris disks \citep{Lagrange2000,Artymowicz1996}, those with $0.01 < f_{\mathrm{disk}} < 0.1$ as either primordial or transitional disks and those with $f_{\mathrm{disk}} > 0.1$ as primordial disks \citep{Binks2016}. Thus, our candidates are most likely primordial or transitional disks, based on the fact that their fractional luminosities $f_{\mathrm{disk}}$ are too high compared to what is usually accepted for debris disks. 

In our sample, only J0501--4337 comes close to this 0.01 debris disk limit, with $f_{\mathrm{disk}}=0.021$. Its estimated disk temperature $T_{\mathrm{disk}}=170 \pm 10\,$K is on the warmer end of the debris disk temperature range ($\sim 30-200\,$K; \citealt{Chen2006,Chen2014,Zuckerman2011}), but the presence of H$\alpha$ emission in its visible spectrum is consistent with the presence of gas in the disk, and its overall SED (see above) brings it closer to the transitional disk status.

%J0501--4337 and J0226--5327  TWA: J0856--1342 and J1247--3816,
J0856--1342 and J1247--3816 have intermediate $f_{\mathrm{disk}}$ values, both at ~0.088, which could indicate either a transition or a primordial disk, in agreement with the overall shape of their SED. Given their estimated age of about $10\pm 3\,$Myr, it is possible for these disks to still be in a primordial state, or at least in a (pre-)transitional state. Generally, stars can retain significant amounts of primordial dust up to $\sim 10\,$Myr, even though they become rarer in the $\sim 8-10\,$Myr range \citep{Rhee2007}. In fact, their incidence is observed to be $\lesssim5\,$\% in TWA and other similar star associations \citep{Williams2011,Hernandez2007,Sicilia2006}. If primordial, or pre-transitional, these disks should also still harbor some gas, which could feed ongoing accretion. Indeed, \cite{Artymowicz1996} proposed that sources with higher fractional luminosities ($f_{\rm disk} > 0.01$), as observed here,  probably contain a significant amount of gas. The presence of a very close-in disk component could explain their high $T_{\mathrm{disk}}$. In fact, the SED for J0856--1342 especially stands out as quite different from the 3 other excess sources as its disk temperature is much warmer. Given the difficulty that we faced in fitting the excess data for this source and the lack of data between the WISE bands 2 and 3, it is difficult at this moment to determine whether the dust is located in an extended disk rather than in one with a central clearing.

\begin{figure}
\centering
    \includegraphics[scale=0.75]{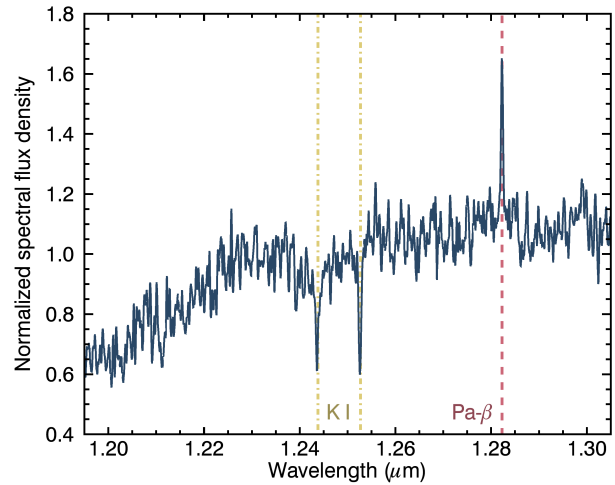}
\caption{K I absorption lines and Pa $\beta$ emission for J02265658--5327032. The Pa $\beta$ emission line is consistent with the presence of a circumstellar disk and a young age. See Section \ref{subsec:youth} for more details.}
\label{fig:PaBeta}
\end{figure}

J0226--5327 has the largest fractional luminosity value in our sample, $f_{\mathrm{disk}} = 0.15$, which is in the typical primordial disk range, although its estimated age of $45\pm 4\,$Myr is well above the expected upper limit for primordial disks around M-dwarfs (e.g. 20\,Myr, \citealt{Binks2016}). The presence of a Paschen $\beta$ emission line in its NIR spectrum, shown in Figure~\ref{fig:PaBeta}, also indicates that this disk indeed contains gas and is consequently primordial or transitional. This is consistent with the overall SED comparison that we obtained. \citeauthor{Binks2016} argue that they found no primordial or transitional disks older than $20\,$Myr in their survey. However, the circumstellar disk dissipation time is known to be smaller for high-mass stars than for solar mass stars \citep{Williams2011}, and they could thus survive even longer for late-type stars and brown dwarfs \citep{Currie2007,Riaz2012}. Given the very low mass of J0226--5327 ($\sim 14\,M_{\rm Jup}$) it may be reasonable for it to still possess a primordial disk for up to 45\,Myr. 
% Binks \& Jeffries, 2016; \emph{in prep., to appear in MNRAS}

In addition, the WISE band $W3$ and $W4$ data indicate that there may be a central clearing in the three disks with cooler dust temperatures, or at least the two that did not need a two-temperature fit. However, the WISE data may just be showing the inner edge of the disk, e.g. the dust could be in an extended disk with a central clearing rather than in a ring. It will be difficult to understand the spatial distribution of the dust without either additional far-IR photometry or spatially resolved images.

The estimated masses of three of our objects are very low, roughly $13-17\,M_{\rm Jup}$, at the lower end of the brown dwarf regime. Other very low-mass substellar objects of similar masses have been previously found to host a disk, for example LOri 156 ($\sim 23\,M_{\rm Jup}$; \citealt{Bayo2012}), OTS 44 ($\sim 12\,M_{\rm Jup}$; \citealt{Luhman2005a,Joergens2013}) and Cha 110913--773444 ($\sim 8\,M_{\rm Jup}$, \citealt{Luhman2005b}), but those have been found in the young $\lambda$ Orionis ($\sim 5\,$Myr; \citealt{Bayo2012}) and Chamaeleon I ($\sim 3$ to $6\,$Myr, depending on the sub-cluster; \citealt{Luhman2007}) star-forming regions, while our targets are probable members of slightly older kinematic associations of young stars ($10-50\,$Myr). 

On a similar note, the brown dwarf TWA~27 (M8) is particularly known for its planetary-mass companion 2MASS~J12073347--393254~b (\citealt{Gizis2002}; Chauvin et al. 2004). \cite{Gizis2002} has shown that the primary brown dwarf features strong H$\alpha$ emission from a near edge-on accretion disk, which was further supported by work from \cite{Mohanty2003} and \cite{GizisBharat2004}. It also displays resolved bipolar jets \citep{Whelan2007}. Our two new detections in TWA, J0856--1342 and J1247--3816, could be analogous to this object, minus the companion.

J0501--4337 and J0226--5327 are most likely the hosts of primordial or (pre-) transitional disks, which would be the first ones at their older age ($\sim45\,$Myr, assuming membership to COL and THA, respectively). If J0501--4337 is a member of $\beta$PMG ($24 \pm 3$\,Myr) it would be more consistent with the presence of primordial gas. These targets could be an indication that low-mass stars and brown dwarfs can keep their gas on a longer timescale than previously suspected. 

%Finally, we are aware that these targets are not representative of the $13\,M_{\rm Jup}$ brown dwarfs sample since they have (or seem to have) a disk. 
It is possible that the presence of disks affected the mass estimates of our targets, which rely on a comparison of their photometry with hot-start models.
%If the models have difficulties to fit these targets, their disks could be the reason why, and their estimated masses could be off.

\section{Concluding remarks}
\label{sec:conclusion}
Synthetic spectra and disk blackbody spectra were fitted to the observed photometric data from the Two-Micron All-Sky Survey (2MASS) and the Wide-field Infrared Survey Explorer (WISE) for a sample of new late-type candidate members of young kinematic stellar associations. This allowed the identification of 4 new objects with infrared excess, indicative of a disk. Disk temperatures and fractional luminosities were determined from this analysis. Their values indicate that the new disks are likely transitional or primordial. The spectral types of these new disk hosts are late (M4.5 to L0$\,\delta$), and correspond to low-mass BDs ($13-19$ and $101-138\,M_{\rm Jup}$) of young ages ($\sim7-13$ and $\sim38-49\,$Myr). These four systems join a relatively small sample of low-mass objects known to harbor a circumstellar disk. There is still a lot to learn about primordial and (pre-)transitional disks around low-mass stars, and these four new candidates could play an important role to shed light on the formation and evolution processes of stars
and planetary systems.

It will be interesting to obtain photometry at longer wavelengths ($>50\,\mu$m) to better constrain the SEDs of these new disks. This would allow verification of multiple-component disks are a better match to the data. Moreover, additional high-resolution spectroscopy for J0856--1342 and J1247--3816 could reveal signs of accretion. These four new disk hosts will be prime targets for polarization studies, high-resolution imaging with ALMA and direct-imaging searches for exoplanets.
%It would also be interesting to detect the polarization signal of light reflected from the disks, or do direct imaging with ALMA.

%Lastly, our candidates are extremely good for exoplanet imaging searches, as the contrast for exoplanet imaging is more favorable for late-type stars and young ages. We may find already formed exoplanet systems or witness their ongoing formation. 

\medskip
\emph{Acknowledgements}

\smallskip
The authors acknowledge financial support for this research from the Fonds de Recherche du Qu\'ebec - Nature et Technologies and the Natural Science and Engineering Research Council of Canada, as well as the Centre de Recherche en Astrophysique du Québec. This work makes use of the data products from the Two Micron All Sky Survey, which is a joint project of the University of Massachusetts and the Infrared Processing and Analysis Center (IPAC)/California Institute of Technology (Caltech), funded by the National Aeronautics and Space Administration (NASA) and the National Science Foundation \citep{2MASS}; the data products from the Wide-field Infrared Survey Explorer, which is a joint project of the University of California, Los Angeles, and the Jet Propulsion Laboratory (JPL)/Caltech, funded by the NASA \citep{WISE}; and the data products from the DENIS $I$-band catalog \citep{DENIScatalog}. This research also made use of the SIMBAD database and VizieR catalogue access tool, operated at Centre de Donn\'ees astronomiques de Strasbourg, France \citep{SIMBAD,VIZIER}; as well as the NASA/IPAC Infrared Science Archive, which is operated by the JPL, Caltech, under contract with NASA. This paper is based on observations obtained at the Canada-France-Hawaii Telescope (CFHT) which is operated by the National Research Council of Canada, the Institut National des Sciences de l'Univers of the Centre National de la Recherche Scientique of France, and the University of Hawaii. It also includes data gathered with the 6.5 meter Magellan Telescopes located at Las Campanas Observatory, Chile, using the FIRE instrument. The authors would like to thank the anonymous referee for their review and suggestions.

%** template de spectre jhk pour les M-L vieille VS low grav?

% -----REFERENCE--------% % % % % % % % % % % % % % % % % % % % % % % % % % % % % %
%\input{autrepartiedutexte.tex}
%\input{Litt_DDR_tableau_2012-09-13.tex}
\bibliographystyle{apj}
\bibliography{ref_all}

% % % % % % % % % % % % % % % % % % % % % % % % % % % % % % % 

\end{document}